\documentclass[sigconf]{acmart}
\usepackage[ruled,vlined,linesnumbered]{algorithm2e}
\usepackage{multicol}
\usepackage{multirow}
\AtBeginDocument{%
  }

\begin{document}

\title{Forgetting Prevention for Cross-regional Fraud Detection with Heterogeneous Trade Graph}


\author{Yujie Li}
\email{yujie\_li@smail.swufe.edu.cn}
\orcid{1234-5678-9012}
\affiliation{
  \institution{Southwestern University of Finance and Economics}
   \city{Chengdu}
   \country{China}
}

\author{Yuxuan Yang}
\email{41810023@smail.swufe.edu.cn}
\affiliation{%
  \institution{Southwestern University of Finance and Economics}
   \city{Chengdu}
   \country{China}
}

\author{Dan Meng}
\email{mengd_t@swufe.edu.cn}
\affiliation{
  \institution{Southwestern University of Finance and Economics}
   \city{Chengdu}
   \country{China}
}

\author{Qiang Gao}
\authornotemark[1]
\email{qianggao@swufe.edu.cn}
\affiliation{
  \institution{Southwestern University of Finance and Economics}
   \city{Chengdu}
   \country{China}
}

\author{Fan Zhou}
\email{fan.zhou@uestc.edu.cn}
\affiliation{
 \institution{University of Electronic Science and Technology of China}
  \city{Chengdu}
  \country{China}
 }

\author{Xin Yang}
\authornote{Corresponding Author.}
\email{yangxin@swufe.edu.cn}
\affiliation{
  \institution{Southwestern University of Finance and Economics}
   \city{Chengdu}
   \country{China}
}

\renewcommand{\shortauthors}{Li et al.}

\begin{abstract}
With the booming growth of e-commerce, detecting financial fraud has become an urgent task to avoid transaction risks. Despite the successful applications of Graph Neural Networks (GNNs) in fraud detection, the existing solutions are only suitable for a narrow scope due to the limitation in data collection. Especially when expanding a business into new territory, e.g., new cities or new countries, developing a totally new model will bring the cost issue and result in forgetting previous knowledge. Moreover, recent works strive to devise GNNs to expose the implicit interactions behind financial transactions. However, most existing GNNs-based solutions concentrate on either homogeneous graphs or decomposing heterogeneous interactions into several homogeneous connections for convenience. To this end, this study proposes a novel solution based on heterogeneous trade graphs, namely HTG-CFD, to prevent knowledge forgetting of cross-regional fraud detection. In particular, the heterogeneous trade graph (HTG) is meticulously constructed from original transaction records to explore the complex semantics among different types of entities and relationships. And motivated by recent continual learning, we present a practical and task-oriented forgetting prevention method to alleviate knowledge forgetting in the context of cross-regional detection. Extensive experiments demonstrate that the proposed HTG-CFD not only promotes the performance in cross-regional scenarios but also significantly contributes to single-regional fraud detection.
\end{abstract}



\maketitle

\section{Introduction}
With the rapid prevalence of digital finance, fraudulent activities always cause huge financial losses, which present a great challenge in reality. As a popular topic in financial applications, fraud detection is essential and urgent for the development of e-commerce/business platforms and has drawn much attention from researchers and practitioners. The main object of fraud detection is to discriminate whether a financial transaction between the customer and a merchant is abnormal. Conventional solutions rely on rule-based models \cite{seeja2014fraudminer} or machine learning-based models \cite{fiore2019using} that resort to constructing hand-craft features stemming from historical trading data to discover the potential anomalous behaviors and dig into the users’ fraud risks. However, the rule-based methods heavily rely on the human prior knowledge, resulting in the detection bias and the collapse of tackling more complex patterns. And most of the existing machine learning-based methods regard fraud detection as a typical binary-classification problem to mine the statistical features of a certain transaction~\cite{shen2007application}, while the interaction between customers and merchants is rarely considered in these solutions.

\begin{figure}[ht]
\centering
 \includegraphics[width =0.45\textwidth]{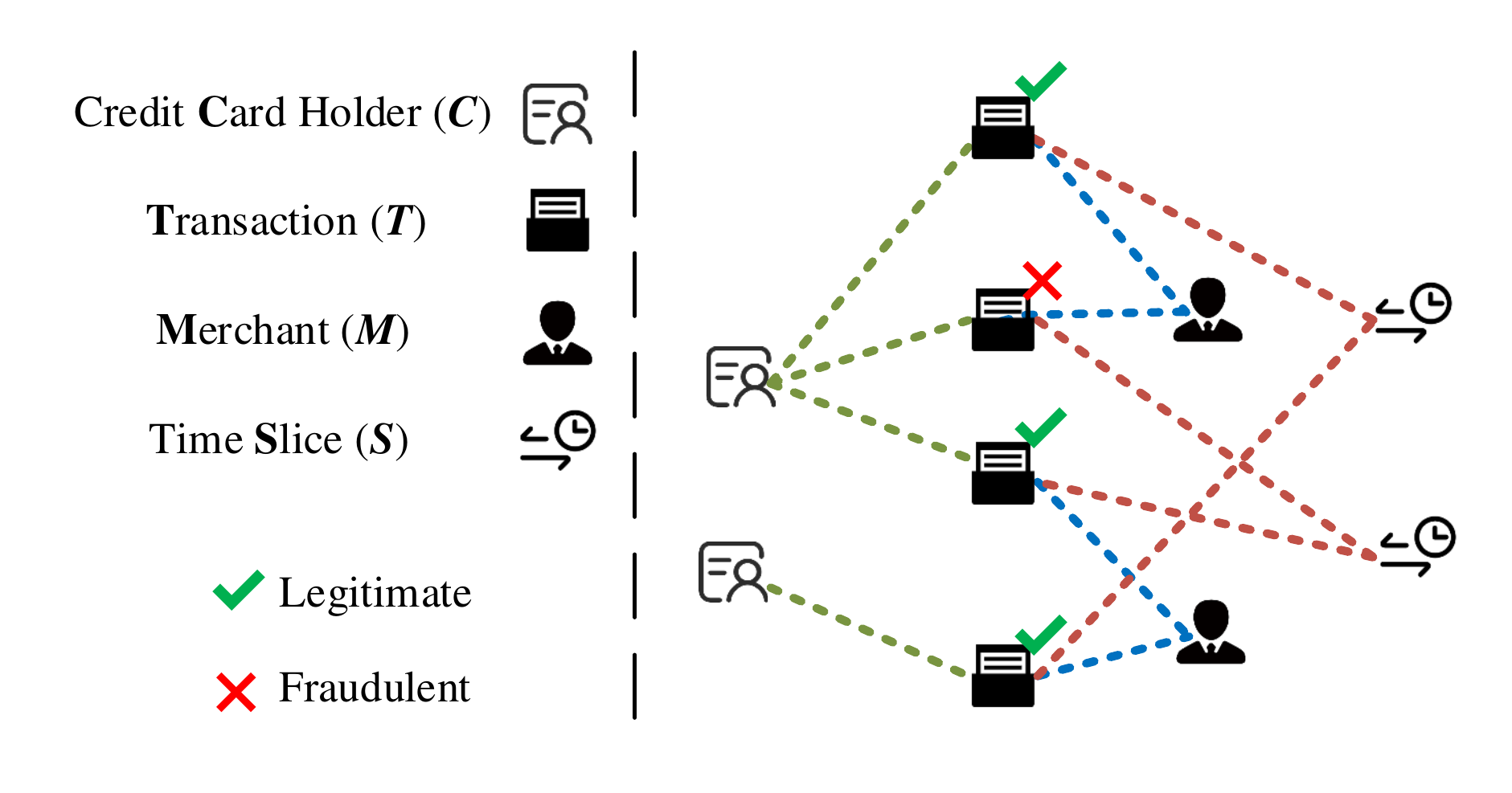}
 \caption{A toy example of a heterogeneous trade graph (HTG). There are four types of entities and three types of relationships. HTG takes transaction nodes as target nodes, so the information from the fraudulent transaction neighbors is easily uncovered during message passing process.}
 \label{fig:toy-example}
\end{figure}

Recently, most endeavors turn to employ graph-based approaches to explore the higher-order and implicit information from trading data by modeling the entities (e.g., customers and merchants) as nodes and the interactions between entities as edges. In particular, the development of deep learning stimulates us to involve graph representation learning to uncover the implied patterns behind the massive financial transactions. For instance, \cite{wang2019fdgars} and \cite{xu2021towards} employ the graph convolution networks to discover the fraudulent users and abnormal applicants of consumer loans by learning multiple relationships among massive entities. To evaluate the distinct impacts of different entities, \cite{wang2019semi} devises a hierarchical attention model to better bridge different neighbors for online fraud detection. However, financial trading data is temporal evolved, and most existing graph-based studies rely on certain entity's features and ignore associated knowledge such as temporal information, which causes omissions of information in learning. \cite{cheng2020graph} presents a spatial-temporal attention-based graph network (STAGN) for credit card fraud detection, which is capable of learning the temporal dynamics of transaction graph features. \cite{liu2021intention} proposes heterogeneous transaction-intention network, which investigates both transaction-intention and transaction-transaction edges.

Nevertheless, we still argue that there are two significant challenges in existing solutions. (1) As shown in Fig.~\ref{fig:toy-example}, real-world transaction data contains various types of attached entities (e.g., trading time) except the customer and merchant entities. To this end, most of the recent GNNs-based solutions concentrate on either homogeneous graphs or simply decomposing heterogeneous interactions into several homogeneous connections, which results in the failure of capturing the higher-order interaction and is unable to investigate the heterogeneous but rich semantic knowledge such as temporal information. (2) In practical applications, existing solutions are only suitable for a narrow scope owing to the limitation of data collection. Specifically, we usually either use the previous trained (static) model or develop a totally new model when expanding business into new territory, e.g., new cities or even new countries, resulting in the uncertainty of capturing financial patterns.
For example, when extending business in a new city, the spatial dynamics are rooted in geographic differences that financial patterns in different regions could be greatly different, whereby existing financial detection model learned from the former regions may not be directly adapted to the current one. To this end, developing a totally new model could be a direct solution to tackle this limitation, however, it will lead to the expensive cost of resources. Instead, re-using the learned model can be a more appropriate solution. 

To remedy the above challenges, this paper investigates a new problem---cross-regional fraud detection (CFD), which aims to continuously detect fraudulent activities in the process of business expansion to new regions. We regard CFD as a continual learning task, and propose a novel solution with \textbf{H}eterogeneous \textbf{T}rade \textbf{G}raph Learning for \textbf{C}ross-regional \textbf{F}raud \textbf{D}etection, entitled \textbf{HTG-CFD}. Firstly, HTG-CFD constructs a comprehensive heterogeneous trade graph (HTG) to expose the complex but semantic interactions among different entities. Next, HTG-CFD builds multiple meta-paths from HTG for diverse semantics learning, whereafter performing with graph-based attention networks to learn the complex structure and interactions among various entities. To address the second challenge, we are inspired by recent continual learning researches and devise a novel task-oriented forgetting prevention module. In contrast to the widely used continual learning solutions, our method contains a prototype-based knowledge replay and a regularized parameter smoothing block to alleviate the forgetting issue for new task learning, where the former can relieve the uncertainty in feature representation and the latter considers the task dependencies during parameter smoothing. The main contributions of this work can be summarized in four aspects:
\begin{itemize}
    \item We investigate the fraud detection problem in a cross-regional context and present a novel solution, HTG-CFD, to tackle the knowledge forgetting throughout learning. To our best knowledge, HTG-CFD is the first work that performs as a continual learning manner for fraud detection.
    \item We propose a comprehensive heterogeneous trade graph (HTG) to integrate temporal semantics with complex interactions among various types of entities and relationships.
    \item We devise a comprehensive but practical forgetting prevention method that, in addition to alleviating the forgetting problem, allows the knowledge consolidation between the new and old tasks without any model expansion.
    \item Our experimental results conducted on real-world datasets demonstrate that HTG-CFD outperforms the baselines in both single-region (static) and cross-regional settings.
\end{itemize}

The remain of this paper is organized as follows. We review the recent studies of fraud detection as well as the advances of graph neural networks and continual learning in Section 2. Then, we make the fundamental definitions and formalize the posed problem in Section 3. Next, we present the details of the proposed HTG-CFD in Section 4. The results of experiments quantifying the proposed method are provided in Section 5. In the end, we conclude this paper and remark future work in Section 6.

\section{Related Work}
We categorize related work into three comprehensive aspects, i.e., fraud detection, graph neural networks, and continual learning. And we also position our work in this context.
\subsection{Fraud Detection}
Fraud detection has always been one of the most vital tasks in the world and has shown its power in preventing detrimental events \cite{ma2021comprehensive}. 
Earlier works mainly used the rule-based methods for fraud detection, assuming that the fraudulent activities have some apparent patterns. Traditionally, the fraud detection problem was studied in various researches by employing satanical machine learning algorithms such as SVM~\cite{mishra2018credit} and Bayesian network~\cite{hajek2017mining}. The fuzzy logic (FL) \cite{supraja2017robust} was employed to design sophisticated models for dealing with complicated financial information. However, these rule-based methods and satanical machine learning algorithms were highly dependent on the human expert knowledge and difficult to find complex and changing patterns \cite{wang2019semi}. With the development of deep neural networks in recent years, DNNs were used effectively to detect financial frauds \cite{ghobadi2016cost,randhawa2018credit,el2017fraud}. Commonly, these methods can extract implicit features from data and thus obtain better detection or prediction results.

Nevertheless, these aforementioned methods fail to regard rich interactions between entities and they always ignore multifaceted information. Transaction data often contain rich correlation information (e.g., temporal interactions), which is not fully exploited by these conventional methods. 

\subsection{Graph Neural Networks}
Technically, our work is greatly related to graph neural networks as they present tremendous power in leveraging comprehensive neighborhood information and fusing structure attributes. For example, GCN \cite{kipf2016semi}, GraphSAGE \cite{hamilton2017inductive} and GAT \cite{velivckovic2018graph} employed convolutional manners, LSTM architecture and self-attention mechanisms to capture and aggregate complex information among nodes, respectively. Nevertheless, all these previous algorithms are only be implemented on homogeneous graphs. Recent studies were extended to manage heterogeneous information networks (HINs). HetGNN \cite{zhang2019heterogeneous} adopted Bi-LSTMs for different node types to aggregate neighbor information. HAN \cite{wang2019heterogeneous} proposed a hierarchical attention aggregation that integrated node-level and semantic-level information from multiple meta-paths.

Recently, as a robust, reliable, and promising detection technique, the graph representation learning was considerably contributed in frauds detection \cite{zhang2021efraudcom,zhang2021fraudre}. In \cite{cheng2020graph,hei2021hawk,jiang2021financial}, different types of objects were considered into building heterogeneous information networks. \cite{pareja2020evolvegcn} explored the dynamics of graph-structured networks. \cite{9338258} modeled dynamic connectivity patterns through bipartite graph embedding, and then detected anomaly from the graph. \cite{qian2021distilling} jointly modeled both structured relations and unstructured information with heterogeneous graph in drug trafficker detection. \cite{wang2021deep} propose a graph construction method for GNN-based fraud detection on the non-attributed graph using a graph pre-training strategy.

Previous studies have shown the ability of GNNs in information aggregation. In this paper, we generate multiple types of meta-paths and entities to capture heterogeneous patterns among transactions. Our proposed model not only consider heterogeneous structure but also explore complex node information with temporal semantics.

\subsection{Continual Learning}
In this work, we formulate and deal with cross-regional fraud detection problem through continual learning. 
Broadly speaking, the existing continual learning solutions can be distinguished into three branches, based on how task-specific information is stored throughout learning: (1) replay-based methods, (2) regularization-based methods, and (3) parameter isolation methods. For replay-based methods, \cite{lao2021two} utilized pseudo rehearsal approaches with no previous samples. For regularization-based methods, \cite{kirkpatrick2017overcoming} mitigated forgetting by penalizing the changes of important weights and \cite{ahn2019uncertainty} designed a neural network based on Bayesian online learning framework with variational inference. For parameter isolation methods, \cite{mallya2018packnet} fixed the learning framework and masked parts of previous tasks to alleviate forgetting. In real-world applications, learning a graph continuously is often necessary. \cite{Han2020GraphNN} employed continual learning to train a GNN incrementally and \cite{zhou2021overcoming} explored continual graph learning (CGL) and presented an experience replay-based framework. Hence, the successful applications of continual learning stimulate us to apply continual learning to HTGs and preserve task-specific knowledge via alleviating catastrophic forgetting. 

To the best of our knowledge, we are the first to bridge the gap between heterogeneous trade graph and continual learning. Specifically, we capture the heterogeneous information from heterogeneous trade graphs, and prevent forgetting in cross-regional fraud detection through continual learning. 

\section{Preliminaries}
We now introduce the definitions and notations we use throughout this paper, followed by the formal definition of the cross-regional fraud detection problem.

Inspired by~\cite{chang2015heterogeneous}, we formally define each heterogeneous trade graph $G_l$ stem from region $\mathcal{R}_l$ as below:
\begin{definition}\label{heterogeneous network}
Let $\mathcal{G}_l=(\mathcal{V}_l,\mathcal{E}_l,\mathbf{X}_l)$ denote a heterogeneous trade graph (HTG) extracted from $l$-th region's transaction data $\mathcal{D}_l$, where $\mathcal{V}_l$ is the node set, $\mathcal{E}_l$ is the edge set and $\mathbf{X}_l$ is the initial feature set. $\mathcal{G}_l$ is associated with a node type mapping function $\phi : \mathcal{V}_l \rightarrow \mathcal{A}$ and an edge type mapping function  $\psi : \mathcal{E}_l \rightarrow \mathcal{O}$, where the $\mathcal{A}$ and $\mathcal{O}$ respectively represent the set of node types and the set of edge types of $\mathcal{G}_l$. Notably, we have $|\mathcal{A}|+|\mathcal{O}| > 2$ since $\mathcal{G}_l$ is a heterogeneous graph. 
\end{definition}
The goal of cross-regional fraud detection is to provide accurate predictions in specific region by node representation learning, which can be formulated as follows:
\paragraph{\textbf{Problem definition.}} \textit{Given a set of heterogeneous trade graphs of different regions $\mathcal{G} = (\mathcal{G}_1, \cdots,\mathcal{G}_l,\cdots, \mathcal{G}_n)$ with $\mathcal{G}_l=(\mathcal{V}_l,\mathcal{E}_l,\mathbf{X}_l)$, we aim to learn a series of neural function $\Phi=(\Phi_1,\cdots,\Phi_l,\cdots,\Phi_n)$, and each function $\Phi_l: \mathcal{V}_l \to \mathbb{R}^d$ maps nodes to a low-dimensional space $d$, i.e., $d \ll |\mathcal{V}_l|$. Hereafter, using a dense layer $f_l: \mathbb{R}^d \to \{0, 1\}$ to detect whether a given node is fraudulent, where 1 refers to fraudulent node and 0 denotes the normal node.}

\begin{table}[!t]
 \begin{center}
  \setlength{\tabcolsep}{1mm}{
  \caption{Notations and explanations.} 
  \begin{tabular}{c c}
\hline
\textbf{Notation} & \textbf{Explanation} \\ \hline
$\mathcal{D}_l$ & The transaction dataset of region $l$\\
$\mathcal{G}_l$ & The heterogeneous trade graph of region $l$\\
$\mathcal{V}_l$ & Node set of $\mathcal{G}_l$\\
$\mathcal{E}_l$ & Edge set of $\mathcal{G}_l$\\
$\mathbf{X}_l$ & Feature set of $\mathcal{G}_l$\\
$\mathcal{P}$ & A meta-path\\
$N^{\mathcal{P}}_{i}$ & Meta-path based neighbors of node $i$\\
$\mathbf{W}$ & A transformation matrix\\
$\mathbf{h}^i$ & Initialized hidden representation of node $i$ \\
$\mathbf{e}^{\mathcal{P}}_{(i,j)}$ & The attention coefficient between node pair $(i,j)$\\
$\mathbf{\alpha}^{\mathcal{P}}_{(i,j)}$ & The normalized attention between node pair $(i,j)$\\
$\textbf{z}^\mathcal{P}_i$ & Meta-path based representation of target node $i$\\
$\textbf{Z}_\mathcal{P}$ & Semantic-specific node embedding of meta-path $\mathcal{P}$\\
$\beta_{\mathcal{P}}$ & The importance of meta-path $\mathcal{P}$\\
$\textbf{Z}$ & The final representation\\
$\mathbb{B}_l$ & The experience buffer form  $\mathcal{D}_l$\\
$\mathbb{F}$  & The Fisher Information matrix\\
\hline
\end{tabular}}
\label{tab: notations}
 \end{center}
\end{table}

\begin{figure*}[ht]
\centering
\includegraphics[width=0.85\textwidth]{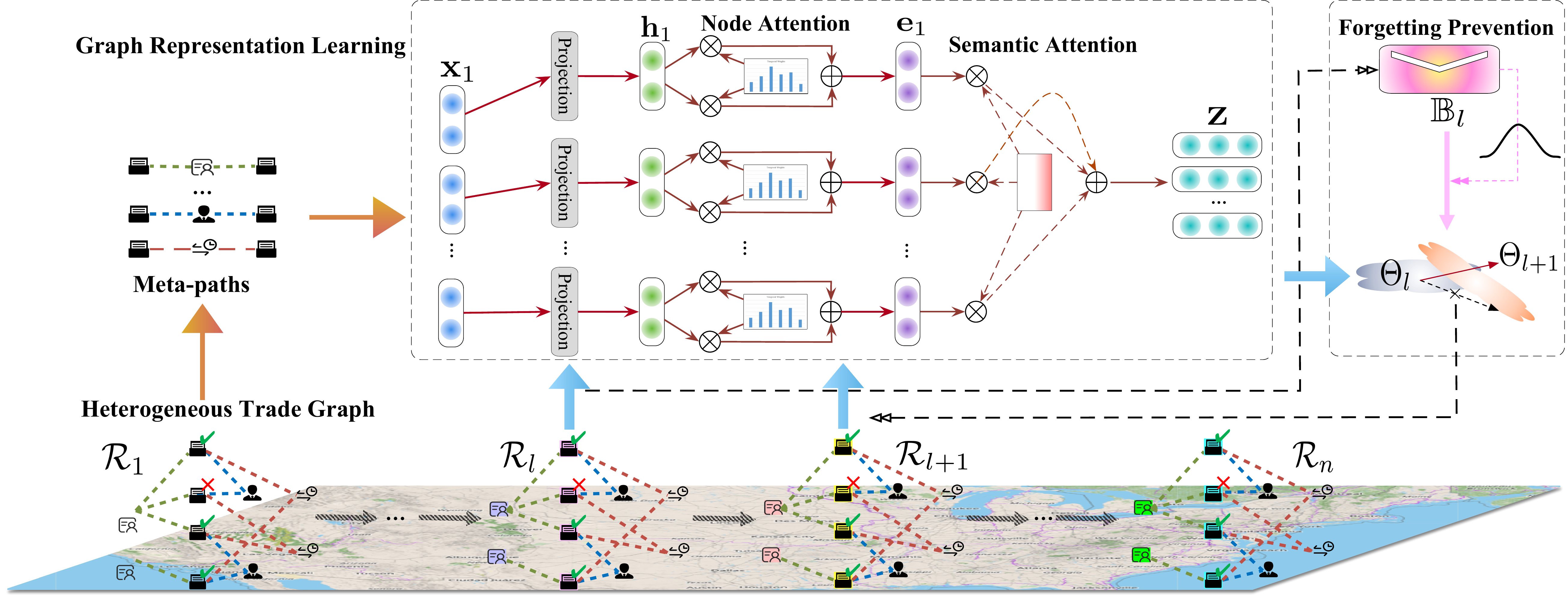}
\caption{The framework of HTG-CFD.}
\label{framework}
\end{figure*}
\section{Methodology}
We turn to explain the specifics of HTG-CFD. We first provide an overview of HTG-CFD and then discuss the technical details of each component. Algorithmic aspects will be introduced at the end.
\subsection{Overview}
We aim at distilling informative knowledge of heterogeneous trade graph (HTG) from a sequence of regions and promising cross-regional performance via forgetting prevention. The main pipeline of our proposed HTG-CFD is shown in Figure \ref{framework}. Firstly, the HTG-CFD contrives a HTG from the current region $l$. Then, HTG-CFD employs a graph representation learning module with attention mechanisms to generate a latent vector for each node, which can be further regarded as the input of the binary classifier for fraud detection. When a new transaction data from region $l+1$ is coming, we use the currently trained model to update the coming HGT stem from region $l+1$, where the complete transaction data from region $l$ cannot be acquired. To consolidate previous knowledge, we use two forgetting prevention strategies, i.e., knowledge replay and parameter smoothing. We will elaborate on the details in the following.

\subsection{Heterogeneous Trade Graph Contriving}
\begin{figure}[ht]
\vspace{-0.4cm}
\centering
\includegraphics[width=0.5\textwidth]{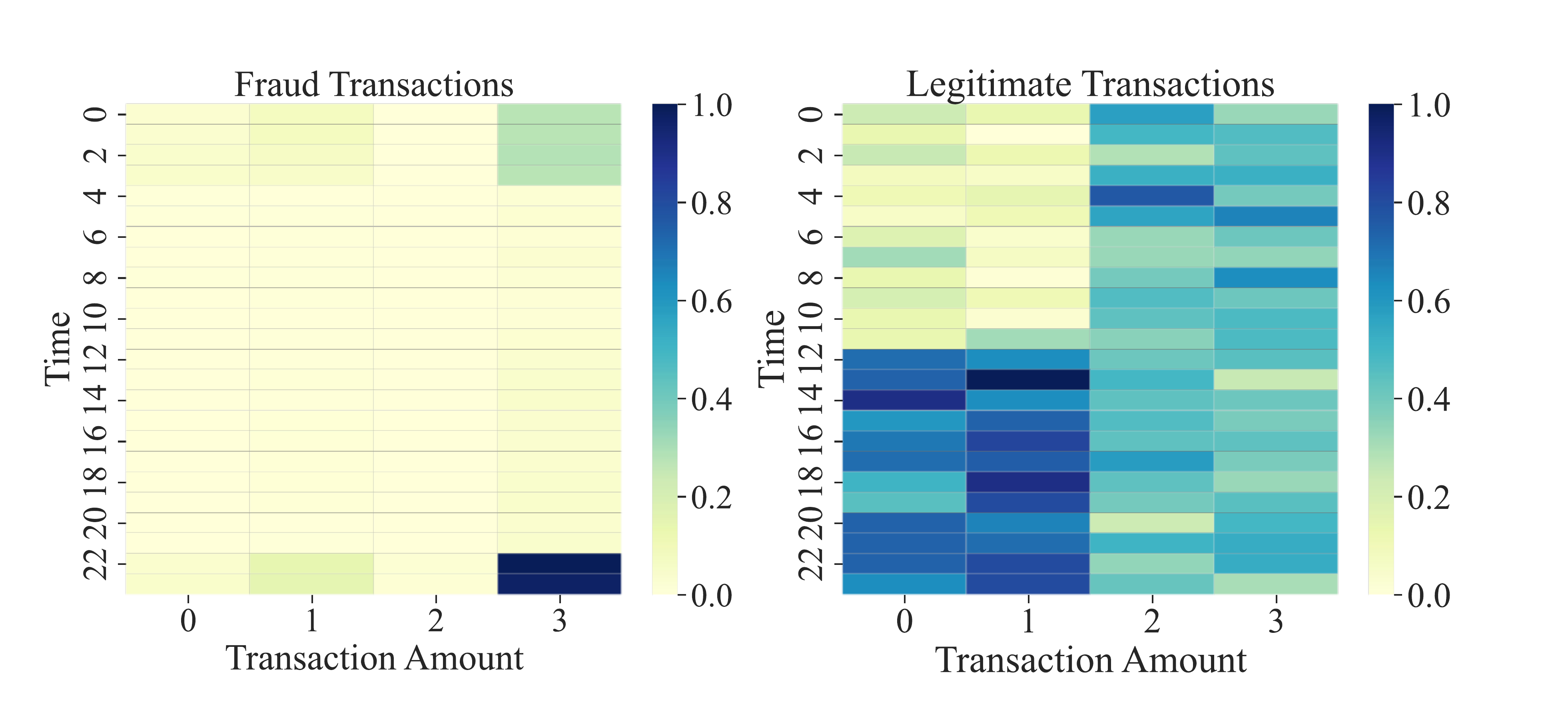}
\caption{Heat maps of temporal slices.}
\label{temporal}
\vspace{-0.2cm}
\end{figure}

As a prerequisite, we first construct our HTG from original transaction data to cater to graph representation learning. There are four types of nodes in our HGT $G_l$, i.e., credit card holders/customers $C_l$, merchants $M_l$, time slices $S_l$, and transaction IDs $T_l$. That is to say, $\mathcal{V}_l={\{ C_l, M_l, S_l, T_l \}}$ and $|\mathcal{A}_l|=4$. Notably, we use the timestamp as a node type when constructing graph-structured data. 
From Figure \ref{temporal}, we find that the frequency of fraudulent transactions changes abruptly across different time periods while legitimate transactions are much more stable. To alleviate the scale of timestamps, we split the time slices into 24 fixed intervals, e.g., [0:00, 1:00] and [1:00, 2:00].

\paragraph{Meta-path.} It is widely used in semantics exploration, aiming at connecting a pair of nodes with a composite relationship such as Deepwalk~\cite{perozzi2014deepwalk} and HAN~\cite{wang2019heterogeneous}. Such a paradigm motivates us to generate massive meta-paths from HTG to explore the diverse semantic interactions among heterogeneous entities. For example, \textit{transaction-card holder-transaction} depicts the co-customer relation while \textit{transaction-time slice-transaction} describes the similar temporal relationship between different transactions. Formally, we define a meta-path $\mathcal{P}$ as $\mathcal{V}^{1} \stackrel{R_1}{\longrightarrow} \mathcal{V}^2 \stackrel{R_2}{\longrightarrow} \cdots \stackrel{R_k}{\longrightarrow}  \mathcal{V}^{k+1}$, where $\mathcal{V}^1$ to $\mathcal{V}^{k+1} $ are $(
k+1)$ different types of nodes. The relations between node types $\mathcal{V}^1$ to $\mathcal{V}^{k+1}$ are denoted as a composite relation $R = R_1 \circ R_2 \circ \cdots R_k$, where the $\circ$ is the composition operator. In the end, we enumerate the existing relationships among target nodes as the predefined meta-paths and concentrate on three types of meta-paths, i.e., TCT (transaction-card holder-transaction), TMT (transaction-merchant-transaction) and TST (transaction-time slice-transaction). We note that most of the previous efforts only regard the temporal information as the node feature, but we argue the temporal information is especially crucial in financial fraud detection. Thus, we regard it as one of the node types as well.

\paragraph{Node feature.}
Beyond the basic heterogeneous graph structure, there exist several informative attributes with respect to each transaction, such as trading location, amount, category. And these attributes depict the inherent semantics of each transaction, which could boost the downstream fraud detection task. Thus, we regard them as the initial node features for the following graph representation learning. Given $ \{C_l, M_l, S_l, T_l \}$, each transaction node $v_l^i \in \mathcal{V}_l$ is associated with a feature vector $\mathbf{x}_l^i$. In the end, our HGT is not only depicted as heterogeneous graphs with multiple types of nodes and relations but also associated with different feature spaces.

\subsection{Graph Representation Learning}
As different types of nodes hold different semantic spaces, we firstly transfer each target node into a new unified semantic space. Given a node $\mathbf{x}_l^i$, it is the initial representation of a target node $i$ in $\mathcal{G}_l$, and we define a transformation matrix $\mathbf{W}_{x}$ to map the initial $\mathbf{x}_l^i$ to a new feature space:
\begin{equation}
 \mathbf{h}_l^i= \mathbf{W}_{x} \cdot  \mathbf{x}_l^i ,
\end{equation}
where $\mathbf{h}_l^i$ is the transformed representation of node $i$ in a new semantic space. Notably, we omit $l$ in the following parts for simplicity.

After projecting all the initial representations into a new latent space, we then leverage a node-oriented attention mechanism to automatically learn the weights among the target node and its neighbors. Following recent attention mechanism in \cite{velivckovic2018graph}, the attention between a node pair $(i,j)$ is the importance of the target node $i$ for its meta-path based neighbor node $j$. Given a pair of node $(i,j)$ which connected via a meta-path $\mathcal{P}$, the attention coefficient $\mathbf{e}^{\mathcal{P}}_{(i,j)}$ can be formulated as follows:

\begin{equation}
 \mathbf{e}^{\mathcal{P}}_{(i,j)} = att_\eta(\mathbf{h}^i,\mathbf{h}^j;\mathcal{P}),
\end{equation}
where the $\mathbf{e}^{\mathcal{P}}_{(i,j)}$ is asymmetric and the $att_\eta$ is an attention mechanism through the deep neural networks parameterized by $\eta$. Then, we perform normalization to get the weight coefficient $\mathbf{\alpha}^{\mathcal{P}}_{(i,j)}$ using $Softmax$ function: 

\begin{align}
\begin{split}
&  [ \mathbf{\alpha}^{\mathcal{P}}_{(i,1)}|| \mathbf{\alpha}^{\mathcal{P}}_{(i,2)} || ... || \mathbf{\alpha}^{\mathcal{P}}_{(i,j)}||... ] \\
& = Softmax([ \mathbf{e}^{\mathcal{P}}_{(i,1)} || \mathbf{e}^{\mathcal{P}}_{(i,2)} || ...|| \mathbf{e}^{\mathcal{P}}_{(i,j)} ||... ]) , j \in N^{\mathcal{P}}_{i}, \\
\end{split}
\end{align}
where $||$ denotes the concatenation operation and $N^{\mathcal{P}}_{i}$ is the set of meta-path based neighbors of target node $i$. Hence, the meta-path based representation of target node $i$ can be obtained from its neighbors with corresponding attention coefficients:
\begin{equation}
    \textbf{z}^\mathcal{P}_i = \sigma(\sum_{j \in N^{\mathcal{P}}_{i}} \mathbf{\alpha}^{\mathcal{P}}_{(i,j)} \cdot \mathbf{h}^j), 
\end{equation}
where $\textbf{z}^\mathcal{P}_i$ is the learned representations of target node $i$. 

Inspired by \cite{wang2019heterogeneous}, we also fuse multiple semantics in a heterogeneous trade graph with multiple-type meta-paths, and learn the importance of different meta-paths to generate the final representations. For a meta-path $\mathcal{P}$, we can obtain a semantic-specific node embedding $\textbf{Z}_\mathcal{P}$. Suppose there are $m$ meta-paths, the learned weights of each meta-path can be formulated as follow:
\begin{equation}
    ( \beta_{\mathcal{P}_1},...,\beta_{\mathcal{P}_m}) = att_\psi(\mathbf{Z}_{\mathcal{P}_1},...,\mathbf{Z}_{\mathcal{P}_m}),
\end{equation}
where $att_\psi$  with learnable parameters $\psi$ denoted the attention operation on semantic-level, according to the meta-path set $\{ \mathcal{P}_1,...,\mathcal{P}_m \}$. 

To training the weights of meta-paths $( \beta_{\mathcal{P}_1},...,\beta_{\mathcal{P}_m})$, a nonlinear transformation is applied to the semantic-specific embedding. By comparing the transformed embedding with a semantic-level attention vector $\mathrm{W_1}$ we measure the importance of the semantic-specific embedding. Moreover, we compute the average of all semantic-specific node embedding that can be interpreted as the importance of each meta-path. Each meta-path importance is calculated as follows:
\begin{equation}
    \mathbf{w}^{\mathcal{P}_j} = \frac{1}{|\mathcal{V}|} \sum_{i \in \mathcal{V}} \mathrm{q}^{\mathrm{T}} \cdot \tanh \left(\mathbf{W'} \cdot \mathbf{z}_{i}^{\mathcal{P}_{j}}+\mathbf{b}\right),
\end{equation}
where $\mathbf{W'}$ is the weight matrix and $\mathbf{b}$ is bias. After obtaining the importance of each meta-path, we normalize them via $Softmax$ function:
\begin{align}
\begin{split}
&  [ \mathbf{\beta}_{\mathcal{P}_1}|| \mathbf{\beta}_{\mathcal{P}_2}|| ... || \mathbf{\beta}_{\mathcal{P}_j}||... ] \\
& = Softmax([ \mathbf{w}^{\mathcal{P}_1}|| \mathbf{w}^{\mathcal{P}_2} || ...|| \mathbf{w}^{\mathcal{P}_j} ||... ]) , \\
\end{split}
\end{align}
where $||$ denotes the concatenation operation and the learned weight can be interpreted as the importance of each meta-path. Then, we finally obtain the representation $\mathbf{Z}$:
\begin{equation}\label{equa:final representation}
    \mathbf{Z} = \sum_{i=1}^{m} \beta_{\mathcal{P}_i} \cdot \mathbf{Z}_{\mathcal{P}_i}.
\end{equation}

By aggregating all semantic-specific embeddings, the final representation is formed and can be applied in fraud detection further. As a typical node classification problem, we use a fully-connected layer to generate the probability of being a fraudulent trade. Actually, we only need to detect the frauds from the node representations of transactions, thus we use the $\mathbf{Z}(T^l)$ with respect to transaction nodes for prediction, namely:
\begin{equation}
\label{Eq:infer}
    \hat{Y}^l=f_l(\mathbf{Z}(T^l);\epsilon),
\end{equation}
where $\hat{Y}^l$ is a set of predicted results w.r.t. transaction nodes. Notably, $f_l$ is the dense layer with sigmoid activation function and $\epsilon$ is the trainable parameters. Now, we summarize the loss function for region $l$. As a classification problem, we use the cross-entropy loss to minimize:
\begin{equation}
    \label{eq:s-loss}
    \mathcal{L}(\Theta_l)=- \frac{1}{|T^l|} \sum_{1}^{|T^l|}( y_i^l\log \hat{y}_i^l+(1-y_i^l)\log \hat{y}_i^l), 
\end{equation}
where $\Theta_l$ is learnable parameters in neural model $\Phi_l$. $\hat{y}_i^l$ is the logic score of each transaction node and $y_i^l$ is the ground-truth label. The following Algorithm~\ref{algotithm1} summarizes the pipeline of obtaining graph representation in our HTG-CFD.

\begin{algorithm}[ht]
\caption{Pipeline of graph representation generation.}
\label{algotithm1}
            \For{each meta-path $\mathcal{P}$}{
                Projection $\mathbf{h} \leftarrow \mathbf{W}_x \cdot  \mathbf{x}$\;
                \For{each node i}{
                    Extract meta-path based neighbors ${N}_{i}^{\mathcal{P}}$\;
                    \For{$j \in {N}_{i}^{\mathcal{P}}$}{
                        Calculate the weight coefficient $\mathbf{\alpha}_{(i,j)}^{\mathcal{P}}$\;
                    }
                    Calculate the single type node embedding  $\textbf{z}^\mathcal{P}_i \leftarrow \sigma(\sum_{j \in N^{\mathcal{P}}_{i}} \mathbf{\alpha}^{\mathcal{P}}_{(i,j)} \cdot \mathbf{h}_j)$\;
                }
                Concatenate from all attention heads\;
            }
            Calculate the weight of the single node type $\beta_{\mathcal{P}}$\;
        Aggregate by $\mathbf{Z} = \sum_{j=1}^{m} \beta_{\mathcal{P}_j} \cdot \mathbf{Z}_{\mathcal{P}_j}$.
\end{algorithm}

\subsection{Forgetting Prevention}
When we have learned a neural model $\Phi_l$ with parameters $\Theta_l$, directly employing $\Phi_l$ to tackle the coming region $l+1$ usually confront the (catastrophic) forgetting issue for previous regions. That is to say, we need to make the current model not only enable learning new knowledge from region $l+1$ but also have a capacity to retain the earlier experiences from region $1,2,\cdots,l$. As such, we do not need to retrain or develop a new model to tackle the task that has been trained before. To this end, we present two simple but efficient strategies to alleviate this problem. Inspired by recent successful cases in continual learning~\cite{zhou2021overcoming,kirkpatrick2017overcoming}, the first strategy is setting an experience buffer $\mathbb{B}_l$ for information replay, which can be co-trained with the new HTG $\mathcal{G}_{l+1}$. The second strategy is forcing the new training model $\Phi_{l+1}$ to remember useful knowledge from previous $\Phi_{l}$. We elaborate on the details as follows:
\paragraph{Prototype-based Knowledge Replay}
In order to preserve the existing knowledge from early region $l$, we can sample a small experience buffer $\mathbb{B}_l$ from transaction data $\mathcal{D}_l$ for knowledge replay, where $|\mathbb{B}_l| \ll |\mathcal{D}_l|$. To this end, we first randomly choose some transaction data in $\mathcal{D}_l$. However, we consider that such a buffer would lead to the training instability issue due to sample bias or feature sparsity. We thus propose a prototype method to alleviate the training instability problem caused by random sampling. As each transaction entity in $\mathbb{B}_l$ is associated with an attribute vector $\mathbf{x}_i^l$, our prototypes are based on the average attribute vector regarding $\{x_i^l\}_1^{|\mathbb{B}_l|}$, which can be defined as:
\begin{equation}
    \label{eq:prototype}
    \mathbf{c}^l=\frac{1}{\mathbb{B}_l}\sum_{\mathbf{x}_i^l \in \mathbb{B}_l} \mathbf{x}_i^l.
\end{equation}

We will use $\mathbf{c}^l$ and a Gaussian prior to generate a prototype $\mathbf{x'}_i^l$ for each $\mathbf{x}_i^l$. Finally, we collect a similar buffer $\mathbb{B'}_l$ and use it with $\mathbb{B}_l$ for experience replay together. Notably, we only use the labeled transaction data to enhance the new region learning due to the security issues in the financial application. In our buffer setup, we only sample transactions from the previous task instead of all trained tasks to avoid memory costs.
\paragraph{Regularization-based Parameter Smoothing}
Although the first strategy, to some extent, is capable of alleviating the forgetting issue from the sample consolidation aspect, it is unable to thoroughly familiar with what knowledge extracted from the previous regions. To further alleviate the knowledge forgetting, we are inspired by recent elastic weight consolidation (EWC)~\cite{kirkpatrick2017overcoming}, and first use Fisher Information to evaluate the importance of each parameters in $\Theta_l$ w.r.t. $\Phi_l$, which can be defined as:
\begin{equation}
\mathbb{F}=\frac{1}{|\mathbf{X}^l|} \sum_{x \in \mathbf{X}^l}\left[g\left(x  ; \Theta_l\right) \cdot g\left(x ; \Theta_l\right)^{\top}\right],
\end{equation}
where $g$ is first order derivatives of the loss. Hence, when the new region $l+1$ arrives, we add a smoothing term to constrain the parameter optimization based on $\mathbb{F}_{l+1}$, which can be formulated as:
\begin{equation}
\mathcal{L}_{\text {s}}=\frac{\lambda}{2} \sum_{i} \mathbb{F}_{i+1}\left(\Theta_{l+1}(i)-\Theta_{l}(i)\right)^{2},
\label{smooting}
\end{equation}
where $\lambda$ is a hand-craft weight. Although financial transaction is usually region-related, while fraud patterns have some common characters. For instance, we can observe that fraud have high time correlations as shown in Fig.~\ref{temporal}. Thus, we conjecture that it is necessary to consider the task similarity in the financial context. Inspired by recent multi-task learning~\cite{zhang2020regularize}, we add a regularization term to measure the parameter distance. Hence, we can rewrite Eq.(\ref{smooting}) as follows:
\begin{equation}
\mathcal{L}_{\text {s}'}=\frac{1}{2}\lambda \sum_{i} \mathbb{F}_i\left(\Theta_{l+1}(i)-\Theta_{l}(i)\right)^{2}+\gamma(||( ||\Theta_{l+1}||_2, ||\Theta_{l}||_2 )||_1).
\label{smooting2}
\end{equation}

Herein, the second term is the $l_{2,1}$-norm and $\gamma$ is also a hand-craft weight. In the end, we can summarize our objective for the region $l+1$, that is:
\begin{equation}
    \label{total-loss}
    \centering
    \min_{\Theta_{l+1}}\mathcal{L}(\Theta_{l+1})+\mathcal{L}_{\text {s}'}(\Theta_{l+1}).
\end{equation}

Finally, we present the general workflow of HTG-CFD in Algorithm~\ref{algotithm2}. Given the sequence of cross-regional trade graphs, we firstly initialize the training model in step 1 and prepare an empty experience buffer in step 2. Then, for the first HTG $\mathcal{G}_1$, the graph representation training process follows step 3 to step 8. From step 11 to step 13, we generate a prototype-base knowledge buffer with Gaussian prior to integrate useful information from previous regions into the current $\mathcal{G}_{l+1}$. Next, we update model with regularization-based parameter smoothing from step 14 to step 18. Finally, we obtain the optimal model $\Theta_{l+1}^*$ from current task.

\begin{algorithm}[ht]
\caption{The overall process of HTG-CFD.}
\label{algotithm2}
\KwIn{A sequence of $\mathcal{G} = \{\mathcal{G}_1, \cdots,\mathcal{G}_l,\cdots, \mathcal{G}_n\}$}
Initialize Parameters $\Theta_1$ of HTG-CFD\;
Set $\mathbb{B}=\varnothing $\;
\For{each epoch}{
    Calculate graph representation for $\mathcal{G}_1$ via Algorithm \ref{algotithm1}\;
    Infer node label of $\mathcal{G}_1$ via Eq.(\ref{Eq:infer})\;
    Calculate Cross Entropy Loss via Eq.(\ref{eq:s-loss})\;
    Back propagation and update parameters\;
    }
    Obtain optimal $\Theta_1^*$\;
\For{$l=1,2,\cdots,n-1$}{ 
    Let $\Theta_{l+1}=\Theta_l^*$\;
    Sample a replay buffer $\mathbb{B}_l$ from $G_l$ for experience replay\;
    Generate a twin buffer $\mathbb{B'}_l$ with $c^l$ and Gaussian prior\;
    Add buffers to the current HTG $\mathcal{G}_{l+1}$\;
  
    \For{each epoch}{
        Calculate graph representation via Algorithm \ref{algotithm1}\;
        Infer node label via Eq.(\ref{Eq:infer})\;
        Calculate Loss via Eq.(\ref{total-loss})\;
        Back propagation and update parameters\;
    }
  Obtain Optimal $\Theta_{l+1}^*$\;
}
\KwOut{$\{ \Theta_n^* \}$.}
\end{algorithm}

\section{Experiments}
In this section, we turn to show the details of our experimental evaluations for the purpose of validating the performance of the proposed HTG-CFD compared with the state-of-the-art baselines, with the aim of answering the following research questions.
\begin{itemize}
    \item \textbf{RQ1:} Does HTG-CFD outperform the related SOTA fraud detection models in both single-regional detection and cross-regional detection?
    \item \textbf{RQ2:} How do the key CL components of HTG-CFD benefit the prediction?
    \item \textbf{RQ3:} Is the heterogeneous trade graph (HTG) constructed in this study interpretable?
    \item \textbf{RQ4:} What is the performance with respect to different hyper-parameter settings?
\end{itemize}

\subsection{Experimental Settings}

\subsubsection{Datasets.}

We conduct several experiments on five regions extracted from a financial database which contains more than one million transactions records in the United States Mainland from the duration of 1st Jan 2019 - 31st Dec 2020~\footnote{https://kaggle.com/kartik2112/fraud-detection}. Distinct from geographical-aware applications,
we consider developing new financial businesses is not strictly divided by cities or states, especially for CFD problem. For instance, the business platforms usually make a pilot in the part of a city whereafter expanding business to new regions. Thus, we do not strictly follow the city-wide strategy but ensure there is no overlapping among the regions. We generate five regions according to different geographical ranges as below and randomly shuffle the sequence as:
$\mathcal{R}_1$ $(30^{\circ}N-40^{\circ}N,95^{\circ}W-100^{\circ}W)$, $\mathcal{R}_2$ $(40^{\circ}N-50^{\circ}N,75^{\circ}W-80^{\circ}W)$, $\mathcal{R}_3$ $(30^{\circ}N-40^{\circ}N,75^{\circ}W-80^{\circ}W)$, $\mathcal{R}_4$ $(40^{\circ}N-50^{\circ}N,95^{\circ}W-100^{\circ}W)$, $\mathcal{R}_5$ $(30^{\circ}N-40^{\circ}N,90^{\circ}W-95^{\circ}W)$.
The details are shown in Table \ref{tab:data statistic}.
\renewcommand\arraystretch{0.8}
\begin{table}[ht]
\small
\centering
    \setlength{\abovecaptionskip}{1pt}
	\setlength{\belowcaptionskip}{1pt}
\caption{The statistics of datasets for five regions.}
\label{tab:data statistic}
\begin{tabular}{lccccccccc}
  \toprule
  Region & $\mathcal{R}_1$ & $\mathcal{R}_2$ & $\mathcal{R}_3$ & $\mathcal{R}_4$ & $\mathcal{R}_5$\\
  \midrule
   \# Transactions & 11,291 & 11,198 & 12,478 & 14,513 & 10,673 \\
   \# Card Holders & 103 & 124 & 126 & 115 & 118 \\
   \# Merchants & 700 & 698 & 698 & 699 & 700 \\
   \# Total Edges & 33,873 & 33,594 & 37,434 & 43,539 & 32,019 \\
  \bottomrule
\end{tabular}
\end{table}

\renewcommand\arraystretch{0.8}
\begin{table*}[ht]
\centering
    \setlength{\abovecaptionskip}{1pt}
	\setlength{\belowcaptionskip}{1pt}
	\caption{Performances on single-regional data.}
\label{tab:single-regional results}
\begin{tabular}{lccccccc}
  \toprule
  Metrics & Training $(\%)$ & CARE-GNN & GAS & GEM  & GATNE & ie-HGCN & HTG-CFD \\
 \midrule
  \multirow{4}{*}{Recall} 
   & $20\%$ & \underline{84.09} & 80.76 & 62.08 & 68.94 & 75.48 & \textbf{89.95} \\
   & $40\%$ & 84.35 & \underline{84.86} & 62.24 & 65.01 & 82.29 & \textbf{86.75} \\
   & $60\%$ & \underline{83.97} & 83.11 & 71.65 & 66.32 & 81.22 & \textbf{89.74} \\
   & $80\%$ & \underline{87.60} & 83.11 & 56.92 & 67.43 & 74.28 & \textbf{88.67} \\
    \cline{2-8}
  \multirow{4}{*}{AUC} 
   & $20\%$ & 83.06 & 81.96 & 82.26 & \underline{88.96} & 87.81 & \textbf{94.71} \\
   & $40\%$ & 82.79 & 82.86 & 79.38 & 89.65 & \underline{90.94} & \textbf{95.25} \\
   & $60\%$ & 82.42 & 81.75 & 80.61 & \underline{88.84} & 85.57 & \textbf{95.73} \\
   & $80\%$ & 84.59 & 83.32 & 77.69 & \underline{89.33} & 87.51 & \textbf{95.15} \\
    \cline{2-8}
  \multirow{4}{*}{F1} 
   & $20\%$ & 78.73 & 78.48 & 74.52 & \underline{80.76} & 79.18 & \textbf{87.85} \\
   & $40\%$ & 79.04 & 79.56 & 74.76 & 80.35 & \underline{84.94} & \textbf{86.52} \\
   & $60\%$ & 78.14 & 78.04 & 77.44 & \underline{80.59} & 80.55 & \textbf{87.76} \\
   & $80\%$ & 80.70 & 79.86 & 72.64 & \underline{81.51} & 80.97 & \textbf{87.39} \\
  \bottomrule
\end{tabular}
\end{table*}

\subsubsection{Baselines.}

We firstly compare the proposed HTG-CFD with several representative fraud detection baselines. Among them, the first three are graph-based fraud detection models the last two are the state-of-the-art heterogeneous GNN models which can extract complex semantics:
\begin{itemize}
    \item \emph{CARE-GNN} \cite{dou2020enhancing}: A GNN-based fraud detection model using reinforcement learning to select informative neighbors to generate representations.
    \item \emph{GAS} \cite{li2019spam}: A GCN-based fraud detection model using both heterogeneous and homogeneous graphs to capture information.
    \item \emph{GEM} \cite{liu2018heterogeneous}: A heterogeneous graph neural network for fraud detection using attention mechanism.
    \item \emph{GATNE} \cite{cen2019representation}: An attributed multiplex heterogeneous network that supports both transductive and inductive learning. 
    \item \emph{ie-HGCN} \cite{yang2021interpretable}: An interpretable heterogeneous graph convolutional network that utilizes a hierarchical aggregation architecture, i.e., HAN~\cite{wang2019heterogeneous}.
\end{itemize}

To verify the effectiveness of our proposed forgetting prevention, we also implement six popular continual learning methods to tackle the CFD problem.
\begin{itemize}
    \item \emph{EWC} \cite{kirkpatrick2017overcoming}: A classical CL framework that protects the important weights of previous tasks to overcoming CF.
    \item \emph{GEM-CL} \cite{tang2020graph}: A model using gradient episodic memory to alleviate forgetting, while allowing beneficial transfer of knowledge to previous tasks.
    \item \emph{MAS} \cite{aljundi2018memory}: A CL method called Memory Aware Synapses (MAS) that computes the importance of parameters in an unsupervised online manner.
    \item \emph{UCL} \cite{ahn2019uncertainty}: An uncertainty-regularized CL method which builds on traditional Bayesian online learning with variational inference.
    \item \emph{DER++} \cite{buzzega2020dark}: An experience replay method that integrates knowledge distillation and regularization in rehearsal.
    \item \emph{HAT} \cite{serra2018overcoming}: A task-based hard attention model that preserves previous tasks’ information without affecting the current task’s learning.
\end{itemize}

Following the learning paradigm of each CL baseline, we note that the first three baselines operate the CL in graph representation learning of HTG, while the last three baselines work with CL by taking the pre-trained but learnable node representations from the original transaction database as input.

\subsubsection{Metrics.}

We note that fraud detection is a typical imbalanced classification problem and the evaluation metrics should have no bias to any class. Thus, we evaluate the proposed HTG-CFD and all the baselines with three widely used metrics, i.e., \textbf{F1}, \textbf{Recall} and \textbf{AUC}. The higher scores of these metrics indicate the better performances of the compared approaches.

\subsubsection{Implementation Details.}

We reproduce the baselines and implement our HTG-CFD with PyTorch library, accelerated by one NIVDIA RTX 3090 GPU. We use Adam as the optimizer with a learning rate of 0.05 and a weight decay of 0.001 as default for comparison. We train 200 epochs for each region with early stopping to accelerate the training process, and randomly sample 15\% of each region's transactions for our experience replay buffer, $\lambda$ is 1.5, and $\gamma$ is 0.00025. Note that the source code of the paper will be released after acceptance.

\subsection{Performance Comparisons (RQ1)}

To verify the performances of the proposed HTG-CFD, we compare our model with different baselines under two different settings: for \textbf{single-regional detection}, we compare HTG-CFD with five state-of-the-art graph-based fraud detection models; for \textbf{cross-regional detection}, we compare HTG-CFD with five CL models with consistent GNN backbone. From the results shown in Table \ref{tab:single-regional results} and Table \ref{tab:cross-regional results}, there are several valuable observations.

\subsubsection{Single-regional Detection}

Firstly, we conduct comparisons with baselines for single-regional (i.e., $\mathcal{R}_1$) detection, evaluated by different proportions of training set---20\%, 40\%, 60\%, 80\%---to verify the performance of the proposed HTG-CFD. Table \ref{tab:single-regional results} reports the average experimental results over 10 times. We find that heterogeneous graph-based models such as GATNE and ie-HGCN achieve promising results compared to traditional homogeneous graph-based models, which demonstrates that integrating various heterogeneous structures is capable of boosting the model performance for semantic learning. 

Particularly, HTG-CFD significantly outperforms all baselines, where HTG-CFD outperforms the best baseline by 2\%-8\%. Different from the existing heterogeneous graph neural networks, our HTG-CFD uncovers the impact of time factor in financial fraud detection task and takes temporal semantic information into account, thus improving the model performances. Furthermore, we evaluate the impact of temporal semantics in the following part of Interpretability Analysis.

\begin{table*}[ht]
\centering
    \setlength{\abovecaptionskip}{1pt}
	\setlength{\belowcaptionskip}{1pt}
	\caption{Performances on cross-regional data.}
\label{tab:cross-regional results}
\setlength{\tabcolsep}{0.75mm}{
\begin{tabular}{lccccccccccccccccccc}
		\toprule
        \multirow{2}{*}{Method} & \multicolumn{3}{c}{$\mathcal{R}_1$} & \multicolumn{3}{c}{$\mathcal{R}_2$} & \multicolumn{3}{c}{$\mathcal{R}_3$} & \multicolumn{3}{c}{$\mathcal{R}_4$} & \multicolumn{3}{c}{$\mathcal{R}_5$} & \multicolumn{3}{c}{Average}\\
        \cmidrule(lr){2-4} \cmidrule(lr){5-7} \cmidrule(lr){8-10} \cmidrule(lr){11-13} \cmidrule(lr){14-16} \cmidrule(lr){17-19} &Recall&AUC&F1&Recall&AUC&F1&Recall&AUC&F1&Recall&AUC&F1&Recall&AUC&F1&Recall&AUC&F1\\
        \midrule
        CARE-GNN&83.73&93.17&79.09&81.11&81.75&77.90&84.33&82.91&78.47&80.52&83.44&79.00&82.59&82.18&78.41&82.45&82.69&78.57\\
        GAS&82.98&82.25&78.61&80.63&80.89&78.35&83.03&82.43&78.39&82.48&82.21&79.30&82.09&80.54&76.92&82.24&81.66&78.32\\
        GEM&63.49&82.82&73.83&73.17&80.64&75.95&74.77&83.38&77.33&64.23&84.07&74.44&74.23&83.25&77.85&69.98&82.83&75.88\\
        HTG-CFD*&81.89&\underline{97.83}&\underline{90.77}&83.10&93.11&\underline{88.66}&80.13&\underline{97.64}&\underline{90.33}&83.71&95.52&\underline{89.84}&\textbf{97.25}&\textbf{99.54}&\textbf{96.65}&\underline{92.63}&\underline{96.72}&\underline{91.25}\\
		\midrule
        EWC&88.58&94.89&87.89&\underline{88.97}&93.66&86.40&89.03&84.62&86.99&\underline{91.37}&94.75&86.75&90.17&94.77&86.95&89.62&94.54&87.00\\
        GEM-CL&81.18&85.99&79.72&78.73&86.26&79.02&79.49&84.22&78.21&76.73&84.56&79.66&74.60&86.54&80.42&78.15&85.51&79.41\\
        MAS&\underline{88.66}&95.15&87.87&88.89&93.78&86.90&\underline{89.10}&94.83&87.87&90.69&95.00&86.68&89.34&94.29&86.70&89.34&94.61&87.20\\
        UCL&72.20&88.00&86.13&71.96&87.98&86.03&71.69&88.45&86.18&73.14&91.04&87.02&69.19&87.41&84.29&71.63&88.57&85.93\\
        DER++&69.84&94.34&86.22&70.16&93.77&86.62&68.76&93.00&85.68&70.57&95.85&86.73&66.87&94.01&84.64&69.24&94.19&85.97\\
        HAT&75.37&95.45&87.85&74.65&\underline{95.84}&87.33&73.01&95.97&87.13&77.24&\underline{96.81}&89.15&71.40&95.40&86.25&74.33&95.89&87.54\\
        \midrule
        HTG-CFD &\textbf{95.51}&\textbf{98.08}&\textbf{92.34}&\textbf{91.43}&\textbf{97.44}&\textbf{92.26}&\textbf{90.67}&\textbf{98.01}&\textbf{93.03}&\textbf{96.69}&\textbf{98.53}&\textbf{93.06}&\underline{92.59}&\underline{98.98}&\underline{94.30}&\textbf{93.38}&\textbf{98.21}&\textbf{92.99}\\
        \bottomrule
\end{tabular}}
\end{table*}

\subsubsection{Cross-regional Detection}

Secondly, we compare our HTG-CFD with six CL baselines for cross-regional fraud detection problem, where randomly splitting each region's transactions into training, validation, and testing sets with the ratio of 60\%:10\%:30\%. 
Following the general continual learning evaluation settings, the training data from each region are fed into the model in a randomly shuffled sequence. For each region $\mathcal{R}_l$, the training data is no longer accessible once it has been used. Then, after all regions' training data are used, the learned model are tested with each regions' testing sets, respectively. Furthermore, we provide a variant of HTG-CFD, i.e., HTG-CFD*, which removes the forgetting prevention part. We use CARE-GNN, GAS, and GEM as the static graph-based fraud detection baselines, where each region will have its own trained model for testing. 

\begin{figure}[ht]
\vspace{-0.6cm}
\centering
 \includegraphics[width =0.5\textwidth]{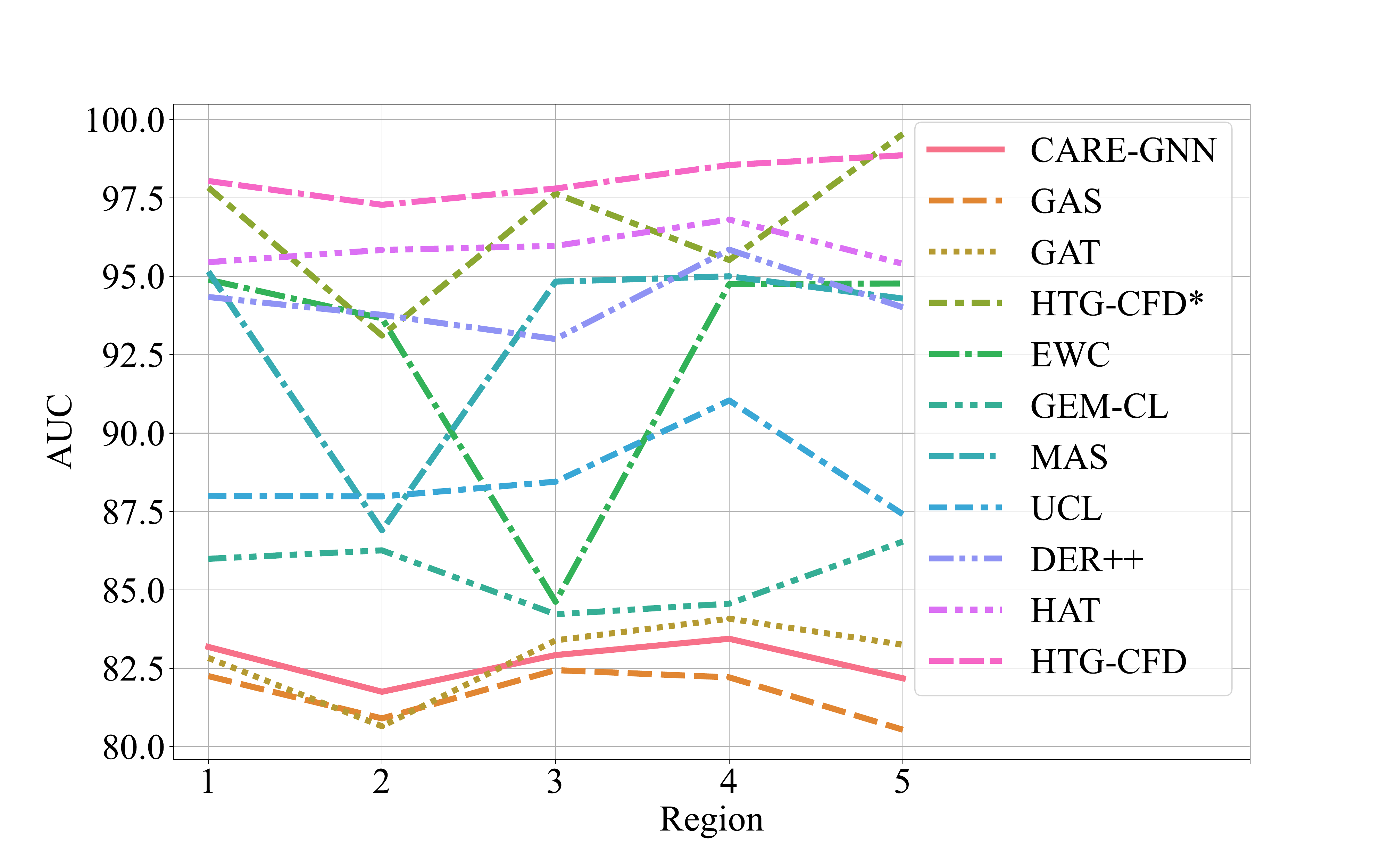}
  \vspace{-0.8cm}
 \caption{AUC variation curve.}
 \label{linplot}
 \vspace{-0.6cm}
\end{figure}

The first four rows of Table \ref{tab:cross-regional results} are the comparisons between HTG-CFD* and static GNN models on CFD problem, the bottom half from Table \ref{tab:cross-regional results} are the comparisons between the HTG-CFD and advanced CL baselines. Table \ref{tab:cross-regional results} shows that our HTG-CFD* and HTG-CFD can achieve comparable performances in cross-regional tasks, which verifies the competitive effectiveness of HTG-CFD and answers the RQ1. Specifically, we have the following discussions:

Compared to the CL baselines, our HTG-CFD achieves the best performances on all metrics, indicating that the HTG-CFD is effective in handling the knowledge forgetting problem of fraud detection.
Especially, HTG-CFD outperforms HAT which has no forgetting problem, demonstrating that HTG-CFD has the promising ability of knowledge transfer that can promote the old tasks' performances. Moreover, existing regularization-based methods (e.g., EWC) and experience replay-based methods (e.g., DER++) perform worse than ours, indicating that directly using off-the-shelf CL methods while ignoring the implicit characteristics behind financial transactions is not conducive to knowledge consolidation and transfer.
In addition to the model accuracy, the observations from Fig.~\ref{linplot} show that our method is more robust in tackling the forgetting problem as the new task incrementally arrives.

Compared to the static models, HTG-CFD outperforms all the baselines from region $\mathcal{R}_1$ to $\mathcal{R}_4$, while HTG-CFD* performs slightly better than HTG-CFD in the last region $\mathcal{R}_5$. The plausible reason is HTG-CFD may sacrifice effectiveness on current task to prevent knowledge forgetting from previous tasks and to guarantee achieving satisfactory averaged performances. We also find HTG-CFD* outperforms all the static graph-based fraud detection baselines from region $\mathcal{R}_1$ to $\mathcal{R}_5$, which further verifies the conclusion in single-regional detection task.

\subsection{Ablation Study (RQ2)}

To well answer the RQ2, we turn to investigate the impact of two key CL components of the proposed HTG-CFD, i.e., Prototype-based Knowledge Replay (PKR) and Regularized Parameter Smoothing (RPS), and verify their effectiveness by removing each part, respectively. We provide two variants of HTG-CFD to study the contributions of PKR and RPS. Specifically, the first variant \textbf{HTG-CFD w/o RPS} removes the RPS module in HTG-CFD, while the second \textbf{HTG-CFD w/o PKR} discards the PKR module. In our implementations, the CL methods, i.e., EWC (parameter smoothing-based) and DER++ (experience replay-based), use the same graph representation learning module as HTG-CFD. Therefore, we again present their test results here for a clear comparison.

Table \ref{tab:ablation results} shows the averaged performance results after all tasks have been trained. 
We find that the proposed HTG-CFD achieves the best performances compared with its two variants on all the metrics. Moreover, the HTG-CFD w/o RPS achieves better performances than DER++ and HTG-CFD w/o PKR, which indicates that capturing knowledge and replaying important knowledge are breakthrough points in continual learning. Besides, the results that HTG-CFD w/o PKR outperforms EWC and HTG-CFD w/o RPS outperforms DER++ demonstrate that either PKR considering the sample bias/uncertainty or RPS considering the task similarity can facilitate the generalization ability of financial transaction learning. 
\renewcommand\arraystretch{0.8}
\begin{table}[h]
\centering
    \setlength{\abovecaptionskip}{1pt}
	\setlength{\belowcaptionskip}{1pt}
	\caption{Ablation Study Results.}
\label{tab:ablation results}
\setlength{\tabcolsep}{1.0mm}{
\begin{tabular}{lccc}
		\toprule
        Method & Recall & AUC & F1\\
        \midrule
        EWC & 89.62 & 94.54 & 87.00\\
        DER++ & 69.24 & 94.19 & 85.97\\
        HTG-CFD w/o RPS & \underline{91.08} & 97.08 & 92.06\\
        HTG-CFD w/o PKR & 89.31 & \underline{97.74} & \underline{92.32}\\
        HTG-CFD & \textbf{93.38}&\textbf{98.21}&\textbf{92.99}\\
        \bottomrule
\end{tabular}}
\end{table}

\subsection{Interpretability Analysis (RQ3)}

To present the interpretability analysis regarding node type selection of our HGT, we randomly mask 0\%, 30\%, 60\%, and 90\% of nodes in HGT for each type. Then, we use Grid Search and SHAP toolkit\footnote{https://pypi.org/project/shap/} to visualize the importance of each type. 

\begin{figure}[ht]
\centering
 \includegraphics[width=0.45\textwidth]{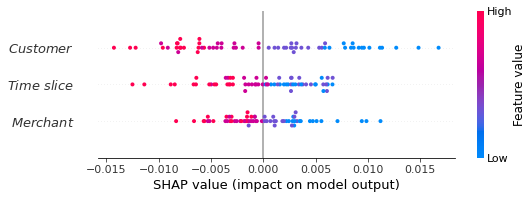}
  \vspace{-0.2cm}
 \caption{SHAP results.}
 \label{Robustness-int}
 \vspace{-0.2cm}
\end{figure}

As shown in Fig.~\ref{Robustness-int}, the more important node type for model training is closer to the top area. Clearly, $Customer$ is most important. This is in line with our natural experience that the occurrence of fraud trade is primarily related to the customers themselves. We also observe that the time slice is more important than merchant, which demonstrates regarding the time slice as a node type is critical for fraud detection. The plausible reason is that some scammers will choose a specific time to defraud. In addition, the point in blue means masking fewer nodes while the point in red means masking more nodes. Obviously, masking more nodes will have a negative impact on model performance since their SHAP value is smaller than 0.

\begin{figure}[ht]
\centering
 \includegraphics[width=0.5\textwidth]{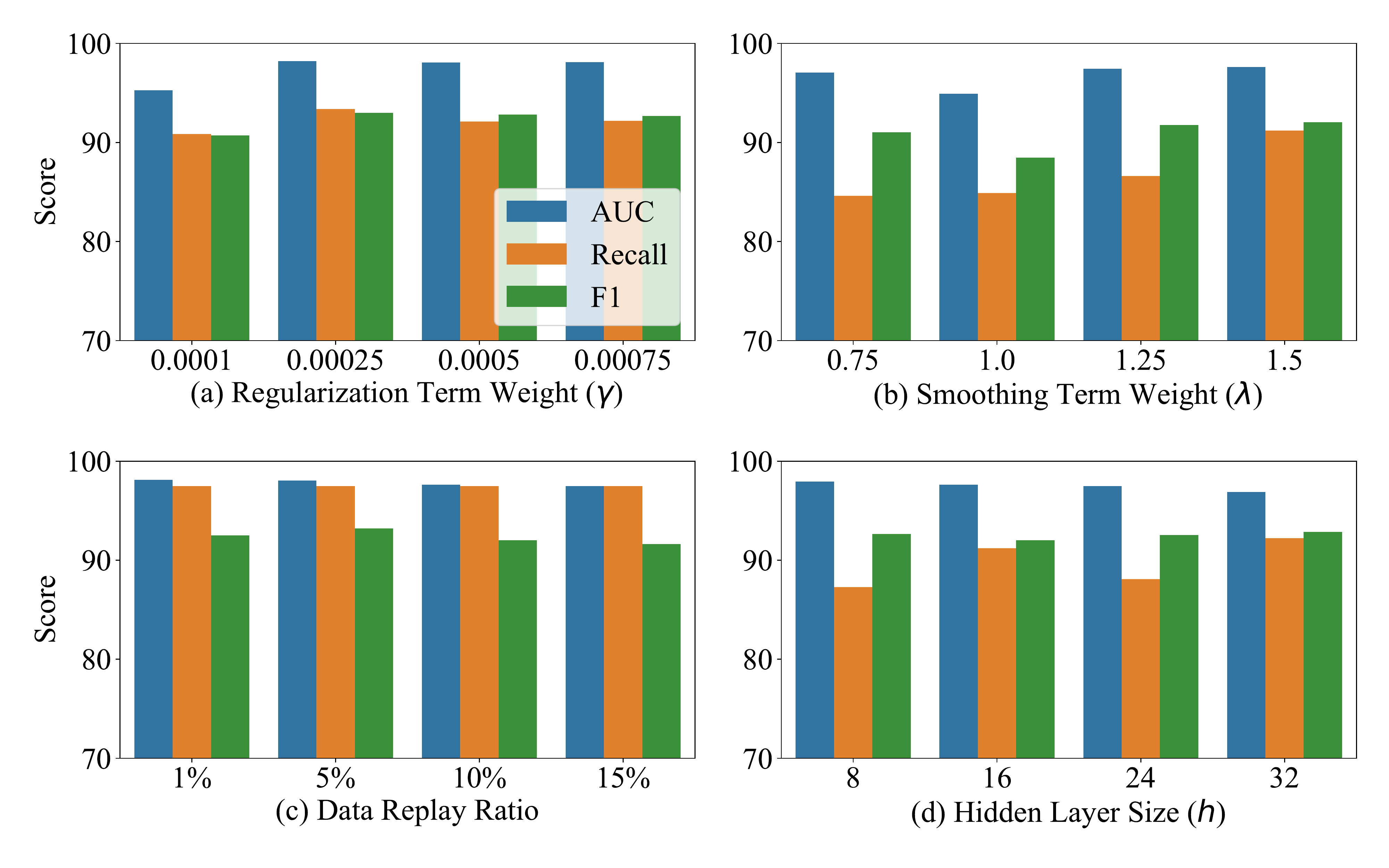}
  \vspace{-0.6cm}
 \caption{Robustness study.}
 \label{Robustness}
   \vspace{-0.4cm}
\end{figure}

\subsection{Robustness Analysis (RQ4)}

To answer the RQ4, we analyze the impact of different parameter settings that could affect the proposed HTG-CFD: $\gamma$, $\lambda$, the hidden size of $|\mathbf{h}|$ and the data replay ratio. 

As shown in Fig~\ref{Robustness}, the proposed HTG-CFD can keep stable performances under different hyper-parameter settings.
Specifically for replay ratio, we vary the percentage of replay ratio from 1\% to 15\% and find that using larger replay buffer will not necessarily bring us higher results, and we safely choose replay ratio as 15\%. Therefore, we can conclude that the proposed HTG-CFD is robust to different hyper-parameter settings.

\section{Conclusions}
In this paper, we construct heterogeneous trade graphs form tabular transaction records to capture more complex semantics. We propose HTG-CFD focusing on cross-regional fraud detection, a graph representation learning model based continual learning strategy. To handle the catastrophic forgetting problem in continual fraud detection, a prototype-based knowledge replay method and a parameter smoothing approach are introduced to achieve forgetting prevention. The proposed HTG-CFD is the first attempt that regards CFD as a continual learning problem, which has the capability of knowledge transfer and relieving expensive resource usage. The experimental results demonstrate that our HTG-CFD outperforms the baselines on both single-regional and cross-regional tasks. 

As for future works, we plan to incorporate more semantic information like social relationships into the model so that the proposed model can uncover more mutual interactions from cross-domain knowledge. Also, we will turn to investigate the expandable network to tackle the cross-regional fraud detection task.

\begin{acks}
This work was supported by National Natural Science Foundation of China (Grant No. 62102326), the Humanity and Social Science Youth Foundation of Ministry of Education of China (Grant No. 20YJC630191), the Key Research and Development Project of Sichuan Province (Grant No. 2022YFG0314) and the Fundamental Research Funds for the Central Universities (Grant Nos. JBK2203010, JBK2207005). We thank the Sichuan Key Laboratory of Financial Intelligence and Financial Engineering for its support.
\end{acks}

\bibliographystyle{ACM-Reference-Format}
\bibliography{bibfile.bib}

\end{document}